\documentclass[showpacs,twocolumn,prl]{revtex4}

\usepackage{graphicx}
\usepackage{graphics}
\usepackage{epsfig}
\usepackage{latexsym}
\usepackage{amsfonts}
\usepackage{amssymb}
\usepackage{amsmath}

\begin{document}
\title{Interaction driven real-space condensation}
\author{M. R. Evans$^{1,3}$, T. Hanney$^{1,3}$, Satya N. Majumdar$^{2,3}$}
\affiliation{$^1$SUPA and School of Physics, University of Edinburgh,
  Mayfield Road, Edinburgh, EH9 3JZ, UK \\
$^2$Laboratoire de Physique Th\'eorique et Mod\`eles Statistiques,
Universit\'e Paris-Sud, Bat 100, 91405, Orsay-Cedex, France\\
$^3$Isaac Newton Institute for Mathematical Sciences, 
20 Clarkson Road, Cambridge CB3 0EH, UK}

\pacs{05.40.-a, 02.50.Ey, 64.60.-i}
\date{\today}

\begin{abstract}
We study real-space condensation in a broad class of stochastic mass
transport models.  We show that the steady state of such models has a
pair-factorised form which generalizes the standard factorized steady
states.  The condensation in this class of models is driven by
interactions which give rise to a spatially extended condensate that
differs fundamentally from the previously studied examples.  We
present numerical results as well as a theoretical analysis of the
condensation transition and show that the criterion for condensation
is related to the binding-unbinding transition of solid-on-solid
interfaces.
\end{abstract}

\maketitle 
Real-space condensation has been observed in a variety of physical
contexts such as cluster aggregation \cite{MKB98}, jamming in traffic
and granular flow \cite{OEC98,CSS00} and granular clustering
\cite{MWL02}. The characteristic feature of these systems is the
stochastic transport of some conserved quantity, to be referred to as
mass; the condensation transition is manifested when above some
critical mass density a single condensate captures a finite fraction
of the mass. The condensate corresponds to a dominant cluster or a
single large jam in these  examples. Perhaps more
surprising realisations of condensation are wealth condensation in
macroeconomies \cite{BJJKNPZ}, where the condensate corresponds to a
single individual or enterprise owning a finite fraction of the
wealth; condensation in growing or rewiring networks where a single
hub captures a finite fraction of the links \cite{DM03} and phase
separation dynamics in one-dimensional driven systems where
condensation corresponds to the emergence of a macroscopic domain of
one phase\cite{KLMST02}.

Mass transport may be modelled in terms of interacting many-particle
systems governed by stochastic dynamical rules. Generically
these systems lack  detailed
balance and  thus have   nontrivial  nonequilibrium steady
states.  Although our understanding of such steady states is still at
an early stage, a class of models has been determined which exhibit a
factorised steady state (FSS) \cite{EMZ04} which can be written as a
product of factors, one factor for each site of the system.  This
simple form for the steady state has afforded an opportunity to study
condensation analytically and has also been used as an approximation
to more complicated nonequilibrium steady states. The conditions under
which condensation can occur have been determined, leading to
conditions on the stochastic mass transport rules for condensation to
result \cite{MEZ05}.  One key feature of the condensate arising in
these models is that it forms at a single site.

In the physical systems of the kind described above, generically the
stochastic transport rules depend not only on the departure site but
also on the surrounding environment.  In general such models do not
have FSS and finding their steady states has remained a challenge.

The purpose of this letter is twofold. First, we introduce a broad
class of mass transport models where the transport rules depend on the
environment of the departure site.  These models do not have an FSS,
yet we can determine their steady states explicitly. The structure of
the steady state generalises the FSS to a pair-factorised steady state
(PFSS).  Secondly, we find that the nature of the condensate in PFSS
is strikingly different from that of the FSS:  unlike in
the FSS, the condensate is spatially extended. This is due to the
short-range correlations inherent in the PFSS, but absent in the
FSS.

We consider a class of mass transport models on a periodic chain with
sites labelled by $i = 1, \ldots, L$. At each site resides a
non-negative integer number, $m_i$, of particles each of unit mass. We
define particle dynamics such that a particle hops from site $i$ to
$i{+}1$ with a rate $u(m_{i{-}1}, m_i, m_{i{+}1})$ (provided $m_i >
0$), so the total mass $\sum_i m_i = M$ is conserved. These dynamics
drive a current of particles through the system.

If the hop rate is only a function, $u(m_i)$, of mass at the departure
site $m_i$, the model reduces to the zero-range process \cite{EH05}
which has a FSS. Explicitly, the probability of a configuration
$\{m_i\}$ occurring in the steady state is
\begin{equation}
P[ \{ m_i \}]
\propto \prod_{i=1}^L f(m_i) \delta\left( \sum_i m_i -M\right)
\label{FSS}
\end{equation}
where $f(m) = 1/ \prod_{k=1}^m u(k)$ for $m\geq 1$ and $f(0)=1$.  Thus
there is one factor $f(m_i)$ for each site $i$ of the system and the
delta function ensures that the total mass is $M$.

When the hop rates $u(m_{i{-}1}, m_i, m_{i{+}1})$ depend on all three
arguments, we propose the PFSS as a natural generalization of the FSS
which takes the following form: the steady state probability of
configuration, $\{ m_i \}$, is
\begin{equation} \label{P(C)}
P [ \{ m_i \} ] = Z_{L,M}^{-1} \prod_{i=1}^L g(m_i, m_{i{+}1})
\,\delta(\sum_{i=1}^L m_i - M) \;.
\end{equation}
Thus there is one factor $g(m_i, m_{i{+}1})$ for each pair of
neighbouring sites.  The normalisation, $Z_{L,M}$, which plays a role
analogous to the canonical partition function in equilibrium
statistical mechanics, is given by
\begin{equation}
Z_{L,M} = \sum_{\{m_i\}} \prod_{i=1}^L g(m_i, m_{i{+}1})
\,\delta(\sum_{i=1}^L m_i - M)\;.
\end{equation}
Note that in the case $g(m_i, m_{i+1})= f(m_i)$, for example, the PFSS
Eq.~(\ref{P(C)}) reduces to the FSS form (\ref{FSS}).

We first establish that the steady state (\ref{P(C)}) holds for a
broad class of mass transport models.  We find that if (though not
only if) the hop rates out of site $i$ factorize~\cite{tbp}:
\begin{equation} \label{factorised}
u(m_{i{-}1}, m_i, m_{i{+}1}) = \alpha(m_{i{-}1}, m_i) \beta(m_i,
m_{i{+}1})\;,
\end{equation}
then the steady state is of the  PFSS form (\ref{P(C)}) with
\begin{equation} \label{g}
g(m, n) = \prod_{i=1}^n \alpha(m, i)^{-1} \prod_{j=1}^m \beta(j,
0)^{-1}\;,
\end{equation}
for $m, n>0$ where $g(0, 0)=1$,
provided $\alpha$ and $\beta$ satisfy the constraint
\begin{equation} \label{constraint}
\frac{\alpha(m{-}1, n)}{\alpha(m, n)} = \frac{\beta(m,
  n{-}1)}{\beta(m, n)}\;.
\end{equation}
Furthermore given  any form of the weight $g(m,n)$
one can determine the functions $\alpha$ and $\beta$ through
the following recursions
\begin{equation}
\label{recursions}
\alpha(l, m) =  \frac{g(l, m{-}1)}{g(l, m)}\;, \quad \beta(m, n)=
\frac{g(m-1, n)}{g(m, n)}\;.
\end{equation}
Thus for
{\em every} choice of $g(m,n)$ there exists a stochastic mass transport
model which will generate the corresponding PFSS.

We now focus on  a particular model which has  a PFSS
with $g(m,n)$ given by
\begin{equation} \label{g(m,n)}
g(m, n) = \exp \left[ -  J | m - n | + \frac{1}{2} U_0 \left(\delta_{m,0}
+\delta_{n,0}\right)\right]\;.
\end{equation}
One can check from (\ref{recursions},\ref{factorised}) that the
corresponding hop rates are
\begin{eqnarray} \label{hop rates}
u(l,m,n) = \left\{ \begin{array}{ll} \exp \left[ -2 J +U_0\delta_{m,1}\right]\;,\quad & {\rm for} \quad m \leq l,n \\  \exp\left[ 2 J +U_0\delta_{m,1}\right]\;,\quad & {\rm for}
\quad m > l,n \\  \exp\left[U_0\delta_{m,1}\right]\;.\quad & \quad
{\rm otherwise}
\end{array} \right.
\end{eqnarray}
Physically, the rate is low if the mass at the departure site is less
than the neighboring masses and is high if the mass is larger than the
neighboring masses. This tends to flatten the density profile and
generates the effective surface tension $J$ in (\ref{g(m,n)}) implying
short-range correlations between the sites. In addition, isolated
particles tend to hop relatively quickly leading to a preference in
the steady state weights for vacant sites. This is reflected by the
on-site attractive potential $-U_0 \delta_{m_i,0}$ in (\ref{g(m,n)}).

The model defined by the hop rates (\ref{hop rates}) is guaranteed to
have a PFSS with $g(m,n)$ in (\ref{g(m,n)}). To investigate whether
the model allows for a condensation transition as the parameters $J,
U_0$ and the conserved mass density $\rho=M/L$ are varied, we have run
Monte Carlo simulations, according to the following prescription. The
system is prepared in a random, homogeneous initial condition and
evolves under random sequential update. During each time step $\delta
t$ a site $l$ is selected randomly and if a particle is present it is
transferred to the neighbouring site $l{+}1$ with probability
$u(m_{l{-}1},m_l,m_{l{+}1}) \, \delta t$. $L$ such time steps
constitute a single Monte Carlo step.
\begin{figure}
\begin{center}
\includegraphics[scale=0.25]{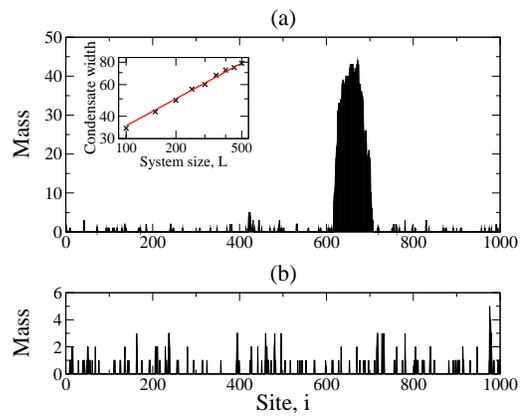}
\caption{Typical steady state configurations for $L=1000$ and $ J =
U_0=1$ (for which $\rho_c = 0.51$) in (a) condensed phase, $\rho = 3$,
and (b) fluid phase $\rho = 1/4$.  The inset in (a) shows an $L^{1/2}$
dependence (indicated by the solid line) of the condensate width on
system size.}
\label{typical config}
\end{center}
\end{figure}

We find that two phases emerge in the steady state depending on $\rho$
and $J$ (where we set $U_0=1$). As illustrated in Fig.\ref{typical
config}, at low density the system resides in a fluid phase, in which
particles are distributed homogeneously throughout the system. When
the density exceeds a critical value $\rho_c(J)$, the system is in a
condensed phase wherein a condensate containing the excess mass
$(\rho-\rho_c)L$ coexists with a critical background fluid of mass
$\rho_c L$. In contrast to a usual condensate that occupies a single
site, as for example in an FSS, the condensate here extends over many
sites. In fact, the
condensate extends over typically $O(L^{1/2})$ sites as shown in 
Fig.\ref{typical config} (a).

\begin{figure}
\begin{center}
\includegraphics[scale=0.25]{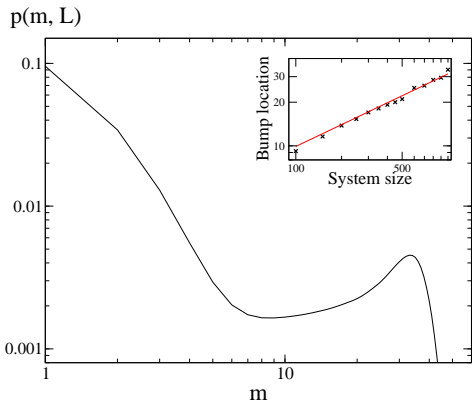}
\caption{Single-site distribution $p(m,L)$ for a system of $L=1000$ sites with
  $\rho = 3$ and $J = U_0 = 1$. The inset shows the
  mass at the maximum of the condensate bump as a function
  of system size $L$, where the crosses are obtained from Monte Carlo
  simulation and the solid line shows an $L^{1/2}$ dependence.}
\label{condensate bump}
\end{center}
\end{figure}

\begin{figure}
\begin{center}
\includegraphics[scale=0.25]{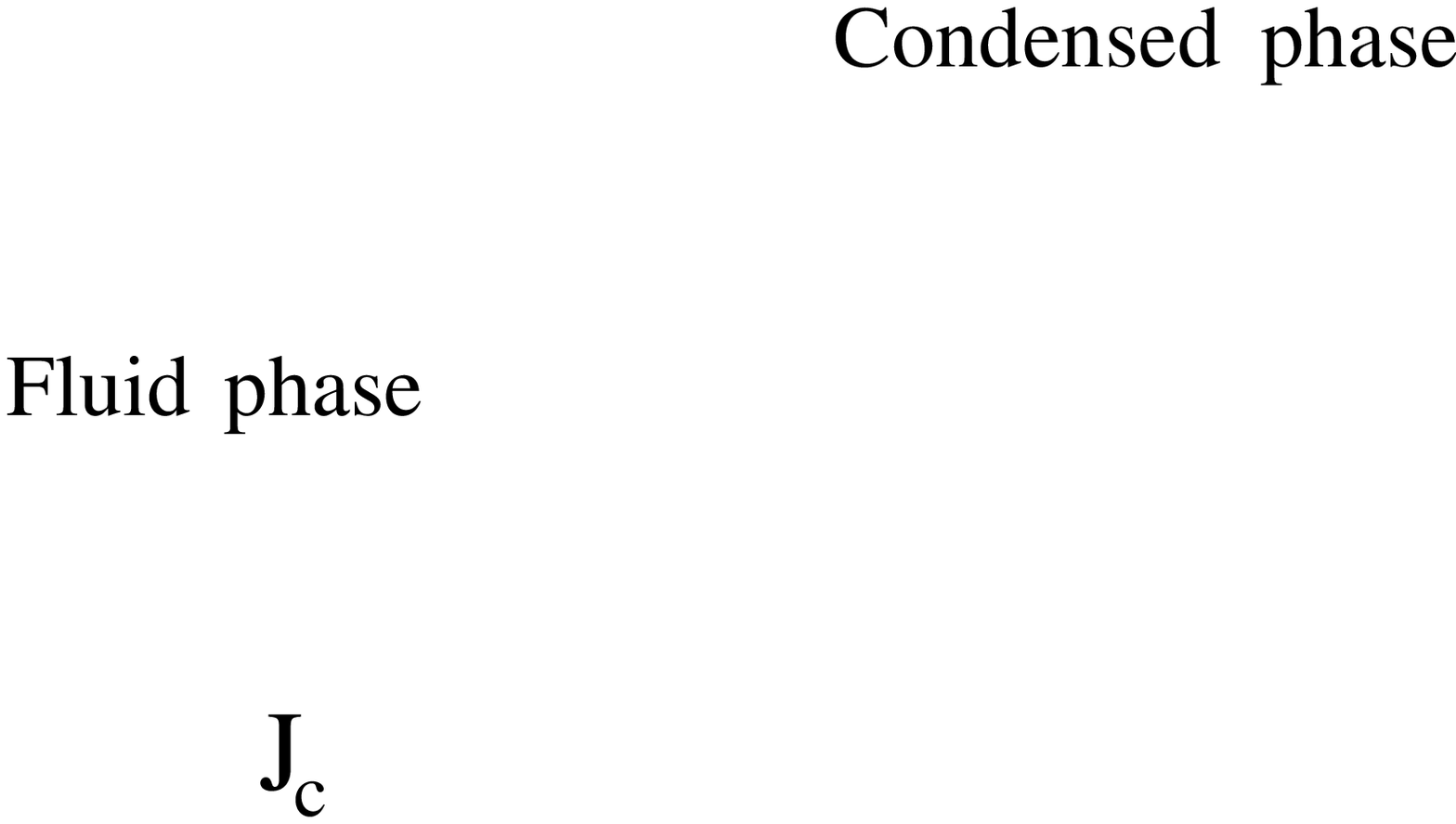}
\caption{Phase diagram for $ U_0 =1$. The crosses are data points
obtained from Monte Carlo simulation. The solid line gives the
theoretical prediction (\ref{explicit rho_c}). For $J<J_c$ given by
(\ref{J_c}) condensation does not occur at any density.}
\label{phase diagram}
\end{center}
\end{figure}

To locate the phase boundary in the $\rho$--$J$ plane we computed the
single-site probabilities $p(m,L)$ that a site contains exactly mass
$m$ in the steady state. In the fluid phase $p(m,L)$ decays
exponentially for large $m$ whereas in the condensed phase an
additional bump emerges at the large $m$ tail of $p(m,L)$ as
illustrated in Fig.~\ref{condensate bump}.
The phase boundary in Fig.~\ref{phase diagram} is determined by the
value of $J$, for fixed $\rho$, at which a bump in $p(m,L)$ first
appears as one increases $J$.
Our theoretical prediction for the phase boundary, presented below, is
in excellent agreement with the numerical results.

The condensate bump in $p(m,L)$ also has an interesting scaling
behavior with $L$.  We plot in the inset of Fig. \ref{condensate bump}
the mass at the maximum of the bump as a function of $L$ and find
that it grows as $L^{1/2}$. This implies that a typical site inside
the condensate has mass of order $L^{1/2}$.  On the other hand since
the condensate carries total excess mass $(\rho-\rho_c)L$ this implies
that there are typically of order $L^{1/2}$ sites inside the
condensate i.e. the spatial extent of the condensate is of order
$L^{1/2}$.

We now analyze the general conditions under which the steady state
(\ref{P(C)}) may admit condensation. The grand canonical partition
function, ${\mathcal Z}_L(\mu)$ (the Laplace transform of $Z_{L,M}$
with respect to $M$)
is given by
\begin{equation}\label{gcpf}
{\mathcal Z}_L(\mu) = \sum_{\{m_i\}} {\rm e}^{-\mu \sum_i m_i}
\prod_{i=1}^L g(m_i, 
m_{i{+}1}) \;,
\end{equation}
where the chemical potential  $\mu$ is determined from the condition
that 
\begin{equation} \label{cp}
\rho  = \rho(\mu)
\equiv
-\frac{1}{L} \frac{\partial {\rm ln}\;
{\mathcal Z}_L(\mu)}{\partial \mu}\;,
\end{equation}
Clearly $\rho(\mu)$ is a decreasing function of $\mu$ for $\mu \geq 0$
\cite{footnote}.  If, as $\mu \to 0$, the function $\rho(\mu) \to
\infty$, then a solution $\mu >0$ of (\ref{cp}) exists for any $\rho
>0$. This implies from (\ref{gcpf}) that the single-site mass
distribution decays exponentially for large $m$ signifying a fluid
phase and there is no condensation. On the other hand, if, as $\mu \to
0$, the function $\rho(\mu)$ approaches a finite value $\rho_c$, then
a solution of (\ref{cp}) can only be found for $\rho< \rho_c$ implying
that the fluid phase exists only for $\rho< \rho_c$.  When $\rho$
exceeds the critical particle density $\rho_c$ there is no solution to
(\ref{cp}) implying the onset of condensation wherein the `excess'
mass $(\rho - \rho_c)L$ is carried by the condensate.

Thus to determine if there is condensation one needs to analyze
(\ref{gcpf}) and (\ref{cp}) as $\mu \to 0$. But, for $\mu =0$,
(\ref{gcpf}) is precisely the grand canonical partition function of a
solid-on-solid (SOS) interface model \cite{CW81,vLH81} where the
interface height at site $i$ is equivalent to the mass $m_i$. Since
$m_i \geq 0$ the interface heights are strictly non-negative implying
that the interface grows on a substrate at $m_i=0$.  Thus $\rho(0)$
from (\ref{cp}) corresponds to the average interface height $\langle
m_i \rangle$ in this SOS model. If $\rho(0) =\infty$, i.e. there is no
condensation transition, the interface is unbound since its mean
height is divergent. On the other hand if there is a condensation
transition, in which case $\rho(0) = \rho_c$ is finite, the interface
is bound with a finite mean height $\rho(0)$.  Therefore the criterion
for a condensation transition is that the corresponding interface
should be bound. Moreover the critical density $\rho_c$ is given by
the mean height of the bound interface.

We now present an exact calculation of the phase diagram in
Fig.~\ref{phase diagram}.  The mean height of this SOS interface model
can be easily calculated using a standard transfer matrix formalism.
The partition function (\ref{gcpf}) may be written as ${\mathcal
Z}_L(\mu) = {\rm Tr}\left[T^L(\mu)\right]$ where the elements of the
transfer matrix $T$ are $g(m',m)$.  In the large $L$ limit only the
eigenvector $| \phi_0\rangle$ of $T$ with the largest eigenvalue
$\lambda_0$ contributes.  The eigenvalue equation reads
$\sum_{m=0}^\infty g(m',m) \langle m| \phi_\alpha\rangle =
\lambda_\alpha \langle m'| \phi_\alpha \rangle$.  The eigenvectors are
either extended states $\langle m| \phi_\alpha\rangle \sim {\rm
e}^{ipm}$ or a bound state $\langle m | \phi_\alpha\rangle \sim t^m$
where $|t|<1$ \cite{CW81}.  If the spectrum contains a bound state
then the bound state corresponds to the largest eigenvalue.
Substituting $\langle m | \phi_0\rangle = t^m$ into the eigenvalue
equation for $m' > 0 $ and $m'= 0$ separately yields $t = {\rm e}^{-
J}/(1{-}{\rm e}^{- U_0})$.  For the bound state to exist $|t|<1$ which
implies
\begin{equation} \label{J_c}
J > J_c = U_0 -  {\rm ln}({\rm e}^{ U_0} - 1)\;.
\end{equation}

Therefore, for $J<J_c$ the system will not condense at any finite
density implying $\rho_c=\infty$, whereas for $J>J_c$, the system
condenses above a finite density $\rho_c$.  The density $\rho_c$ is
given by the mean height in the bound state
\begin{equation} \label{rho_c}
\rho_c = \frac{\sum_m m |\langle m | \phi_0\rangle|^2}{
\sum_m  |\langle m | \phi_0\rangle|^2}\;.
\end{equation}
Using the bound state eigenfunction one finds
\begin{equation} \label{explicit rho_c}
\rho_c = \left[ {\rm e}^{2 (J-J_c)} - 1\right]^{-1}\;.
\end{equation}
This prediction is in excellent agreement with the numerical data as
illustrated in Fig. \ref{phase diagram}.

We now discuss the condensation transition in a more general PFSS
where
\begin{equation}\label{geng}
g(m,n) = K( |m-n|) \exp\left[\frac{1}{2}\left( U(m)+U(n)\right)\right]\;.
\end{equation}
Here $K( |m-n|)$ represents the interaction between nearest neighbor
masses and $-U(m)$ is an on-site potential.  If $K(x)$ decays
sufficiently rapidly for large $x$, as in (\ref{g(m,n)}), for
condensation to happen one only requires $U(m)$ to be positive and
localised near $m=0$. In this case the condensation is
interaction-driven and the existence of the condensation transition
corresponds to having a bound interface. In such cases quite
generically the height and width of the condensate are expected to
scale as $L^{1/2}$. This follows from a Brownian excursion argument:
the localized on-site potential plays no role at sites occupied by the
condensate --- in the absence of the potential, the problem can be
related to a random walk problem where the random walker takes
independent steps with length drawn from a distribution
$K(x)$\cite{SM06}. The shape of the condensate is determined by a
single large loop defined by the excursion of the random walker as
shown in Fig.~\ref{typical config}. The probability that the walker
returns to the origin for the first time after $N$ steps scales as
$N^{-3/2}$ for sufficiently rapidly decaying $K(x)$.  So, the average
number of steps until the first return is $\int_1^{L} N^{-3/2} N \; dN
\sim L^{1/2}$ (the upper cut-off, $L$, is determined by the maximum
number of possible steps). This predicts that the spatial extent of
the condensate is ${\mathcal O}(L^{1/2})$. Also, because it is
Brownian, the typical height of the excursion, and therefore that of
the condensate, scales as $L^{1/2}$.  Note that the area under the
excursion, equivalent to the mass contained in the condensate, is
${\mathcal O}(L)$, as it should be.

This interaction-driven condensation is rather different from the type
exhibited in an FSS. There the function $K(x)=1$ and a localized
on-site attractive potential is no longer capable of driving
condensation.  Instead one requires a specific unbounded potential of
the form $-U(m) \sim \gamma \ln m$ for large $m$ \cite{EH05}. Thus in
the FSS condensation is potential-driven.

To summarize, the steady state (\ref{P(C)}) extends the class of
exactly solvable steady states in nonequilibrium statistical
mechanics. The condensed phase which emerges in a PFSS is
fundamentally different from that of an FSS. In a PFSS the
condensation is interaction-driven and moreover the condensate extends
spatially over ${\mathcal O}(L^{1/2})$ sites.  The explicit form of
the single-site mass distribution $p(m,L)$ in the FSS condensed phase
has been determined recently \cite{MEZ05}. It remains a challenge to
compute $p(m,L)$ for the PFSS condensed phase.  It would also be of
interest to study PFSS in higher dimensions.

Finally we note that the FSS has provided insight into number of
issues of nonequilibrium statistical physics. Although we have
focussed here on the issue of condensation, the generalisation to a
PFSS should allow one to address other interesting issues such as the
role of conservation laws \cite{GS03}, disorder \cite{K00},
boundary-induced phenomena \cite{LMS05} and fluctuation theorems
\cite{HRS05}.

T. H. thanks  the EPSRC for support under programme grant  GR/S10377/01.


\begin{thebibliography}{00}
\bibitem{footnote}
Strictly,
if $g(m,n)$ contains a  pure exponential factor e.g. $\exp( a(m +n))$, $Z_L(\mu)$ in (\ref{gcpf})
would exist  for $\mu \geq a$; but this just amounts to  a shift in the zero of the chemical potential.
\end{thebibliography}
\end{document}